\documentclass[letterpaper, 10pt, conference]{ieeeconf}
\IEEEoverridecommandlockouts 
\usepackage{cite}
\usepackage[mathscr]{eucal}
\usepackage{graphicx}
\usepackage{mathrsfs}
\usepackage{mathtools}
\usepackage{tabularx}
\usepackage{tikz}
\usetikzlibrary{positioning}
\usetikzlibrary{calc,through,backgrounds}
\usepackage{fullpage}
\usepackage{multirow}
\usepackage{optidef}
\usetikzlibrary{matrix,chains,positioning,decorations.pathreplacing,arrows}
\usepackage{soul}
\usepackage{synttree}
\usepackage{neuralnetwork}
\usepackage{url}
\usepackage{multicol}
\usepackage[style=base]{caption}
\usepackage{algorithm}
\usepackage[algo2e]{algorithm2e} 
\usepackage{algpseudocode}%
\usepackage{balance}
\usepackage{filecontents}
\usepackage{bbm}
\usepackage{float}
\usepackage{amsmath,amssymb,enumerate,url}
\usepackage{color}

\usepackage{lipsum} 
\usepackage{tabularx}
\usepackage{booktabs}
\usepackage{footnote}
\makesavenoteenv{tabular}
\makesavenoteenv{table}
\usepackage[colorinlistoftodos,textsize=tiny]{todonotes}
\DeclareGraphicsExtensions{.pdf, .jpg, .png, .jpeg}

\usepackage{geometry}
\usepackage{url}
\usepackage{flushend}

\newcount\Comments  
\definecolor{darkgreen}{rgb}{0,0.5,0}
\definecolor{purple}{rgb}{1,0,1}

\newtheorem{definition}{Definition}
\newtheorem{theorem}{Theorem}

\newtheorem{lemma}{Lemma}

\newtheorem{corollary}{Corollary}

\title{\LARGE \textbf{Online Min Cost Circulation for Multi-Object Tracking on Fragments}}

\author{Yanbing Wang\IEEEauthorrefmark{1}\IEEEauthorrefmark{2}
\and Junyi Ji\IEEEauthorrefmark{1}
\and William Barbour\IEEEauthorrefmark{1}
\and Daniel B. Work\IEEEauthorrefmark{1}
\thanks{Corresponding author: Yanbing Wang, yanbing.wang@vanderbilt.edu}
\thanks{\IEEEauthorrefmark{1}Department of Civil and Environmental Engineering, and the Institute for Software Integrated Systems, Vanderbilt University, Nashville, TN, USA 37240.}
}

\begin{document}
\maketitle
\begin{abstract}
Multi-object tracking (MOT) or global data association problem is commonly approached as a minimum-cost-flow or minimum-cost-circulation problem on a graph. While there have been numerous studies aimed at enhancing algorithm efficiency, most of them focus on the batch problem, where all the data must be available simultaneously to construct a static graph. However, with the growing number of applications that generate streaming data, an efficient online algorithm is required to handle the streaming nature of the input. In this paper, we present an online extension of the well-known negative cycle canceling algorithm for solving the multi-object tracking problem with streaming fragmented data. We provide a proof of correctness for the proposed algorithm and demonstrate its efficiency through numerical experiments.
\end{abstract}

\section{Introduction}

The multi-object tracking (MOT) problem plays a pivotal role in modern computer vision-aided cyber-physical systems. These systems rely on intelligent sensing technologies and efficient data processing tools to monitor and control physical infrastructures. However, challenges emerge when dealing with the increasing data size resulting from larger field of views covered by multiple cameras and the continuous stream of incoming data. To address these challenges, effective algorithms capable of handling significant volumes of streaming data are required. This paper focuses on extending a well-known algorithm designed for the MOT problem into an online framework that can process streaming data of any size. The algorithm described in this paper can effectively handle ``fragments," which refer to incomplete tracking of moving objects caused by conservative tracking from the upstream video processing algorithms. Tracking discontinuities can occur due to various factors such as object occlusion and/or misalignment between adjacent cameras.

Various approaches for data association have been proposed, taking into consideration factors such as association criteria, object motion complexity, and computational requirements. These approaches typically differ in their choices of 1) matching cost (referred to as probability, affinity, energy, or confidence) and 2) matching criteria (such as global cost minimization, greedy approach, hierarchical matching, etc.), leading to different problem formulations. The matching cost incorporates kinematic information (e.g., position and velocity) and attribute information (e.g., shape and appearance), while the matching criteria guide the algorithm in solving the data association problem.

Solving the optimal data association assignment is inherently a challenging NP-hard matching problem that requires combinatorial optimization algorithms. However, specific characteristics of the MOT problem, such as the Markov assumption of association cost, can be leveraged to apply polynomial-time algorithms like bipartite matching and min-cost flow solvers. Graph-based formulations offer efficient algorithms for finding global minimum-cost tracking solutions. In these formulations, tracks (or detections) are represented as nodes in a graph, while pairwise matching costs are represented as graph edges. The general data association problem can be viewed as finding the least-cost set cover on the track graph~\cite{chong2012graph}. Several studies \cite{castnnon2011multi,zhang2008global,vyahhi2008tracking} have investigated efficient algorithms related to bipartite matching and min-cost flow. Interested readers are encouraged to refer to a recent survey~\cite{Rakai2022} for further exploration of this topic.

To the best of our knowledge, the majority of previous studies on graph-based MOT approaches have focused on offline methods. In these methods, all detections/fragments must be available in memory to construct a static graph. However, this approach can be a significant disadvantage, especially as more sensing devices provide sequential data that requires continuous monitoring. While a few online methods, such as \cite{lenz2015followme}, operate on a frame-by-frame basis, they still require multiple iterations of updates within each frame. In contrast, our method operates on a fragments (or tracklets) graph. This means that fragments are added to the graph one at a time, resulting in even further reduced computation requirements.

This paper provides the detailed algorithm for solving the online object matching problem presented in~\cite{wang2022automatic}, which serves to post-process trajectory data from the newly established I-24 MOTION project~\cite{gloudemans202324}. The paper is organized as follows: Section~\ref{sec:prelim} presents the preliminary details of the main technique, min-cost-circulation (MCC), which serves as the foundation for solving the MOT problem. Section~\ref{sec:online_ncc} introduces the online negative cycle canceling (NCC) algorithm for solving MOT problems in a streaming input setting. This section also includes a proof of correctness. In Section~\ref{sec:experiments}, we demonstrate the application of the online algorithm through a numerical experiment. Finally, Section~\ref{sec:conclusion} concludes the paper by summarizing the findings.

\section{Preliminary: The Min-Cost-Circulation (MCC) Problem for Tracking}
\label{sec:prelim}
In this section, we outline the problem formulation for MOT as an equivalent problem for finding the minimum-cost circulation (MCC) of a graph. Solving for MCC on a track graph results in trajectory sets that have the highest \textit{maximum a posteriori} (MAP). The problem formulation is explained in literature such as~\cite{zhang2008global, wang2019mussp}, and therefore only highlighted briefly in this section.

\subsection{Problem formulation}
A fragment with index $k$ is denoted as $\phi_k = \{p_1,...,p_n\}$, which consists of a series of positional data ordered by time (frame). Each data point $p_i$ is a vector containing timestamp, $x$ and $y$ position of a fixed point on the bounding box. We are given a set of fragments as input $\Phi = \{\phi_i\}$. A trajectory $\tau_k=\{\phi_{k_1},...,\phi_{k_n}\}$ consists of one or more fragments. A set of such trajectories form a trajectory set hypothesis $T = \{\tau_1, ..., \tau_K\}$. Assuming that fragments are conditionally independent, the fragment association step aims at finding $T^*$, the hypothesis with the highest MAP:
\begin{equation}
\label{eq: MAP}
    \begin{aligned}
    T^* &=\textrm{argmax}_{T}P(T|\Phi)\\
        &=\textrm{argmax}_{T}P(\Phi|T)P(T)\\
        &=\textrm{argmax}_{T}\prod_iP(\phi_i|T)\prod_{\tau_k\in T}P(\tau_k)\\
        & \textrm{s.t. }  \tau_k \cap \tau_l = \emptyset, \ \forall k \neq l,
    \end{aligned}
\end{equation}
with a non-overlapping trajectory constraint, since each fragment can belong to at most one trajectory. The likelihood $P(\phi_i|T) = P(\phi_i) = \beta_i$ indicates the probability that a fragment is a false positive and thus should not be included in the trajectory hypothesis. The prior of a trajectory can be modeled as a Markov chain:
\begin{equation}
    P(\tau_k) = P_{\text{enter}}(\phi_{k_1})\prod_{i=1}^{n-1}P(\phi_{k_{i+1}}|\phi_{k_{i}})P_{\text{exit}}(\phi_{k_n}),
\end{equation}
where $P_{\text{enter}}(\phi_{k_1})$ and $P_{\text{exit}}(\phi_{k_n})$ denote the probabilities that $\phi_{k_1}$ starts the trajectory and $\phi_{k_n}$ ends the trajectory, respectively. Taking the negative logarithm of~\eqref{eq: MAP}, the MAP problem becomes equivalent to the following integer program:

\begin{subequations}
\label{eq:mcf}
\begin{align}
    &\underset{f_i, f_{i,j}, f_i^{en},f_i^{ex}}{\textrm{minimize}} \quad \sum_i c_if_i + \sum_i c_i^{en}f_i^{en} \notag\\
    &\qquad\qquad +\sum_{i,j} c_{i,j} f_{i,j} + \sum_i c_i^{ex}f_i^{ex}\label{eq:obj}\\
    &\ \ \textrm{subject to}  \qquad f_i, f_{i,j}, f_i^{en},f_i^{ex}\in \{0,1\}, \label{eq:c1}\\
    &f_i^{en}+\sum_{j}f_{j,i}=f_i=f_i^{ex}+\sum_{j}f_{i,j}, \label{eq:c2}
\end{align}
\end{subequations}

where
\begin{equation}
\begin{aligned}
    c_i^{en} = -\log P_{\text{enter}}(\phi_i), \ c_i^{ex} = -\log P_{\text{exit}}(\phi_i), \\
    \ c_{i,j}=-\log P(\phi_i|\phi_j), \ c_i = -\log \dfrac{1-\beta_i}{\beta_i}.
\end{aligned}
\end{equation}

The decision variables are binary according to the unit-flow constraint~\eqref{eq:c1}. $f_i$ indicates whether $\phi_i$ should be included in any trajectory, $f_i^{en}$ and $f_i^{ex}$ determine whether a trajectory starts or ends with $\phi_i$, respectively. $f_{i,j}$ indicates if fragment $\phi_j$ is an immediate successor of $\phi_i$.
The flow-conservation constraint~\eqref{eq:c2} ensures non-overlapping trajectories. 


\begin{figure*}
    \centering
    \includegraphics[width = \linewidth]{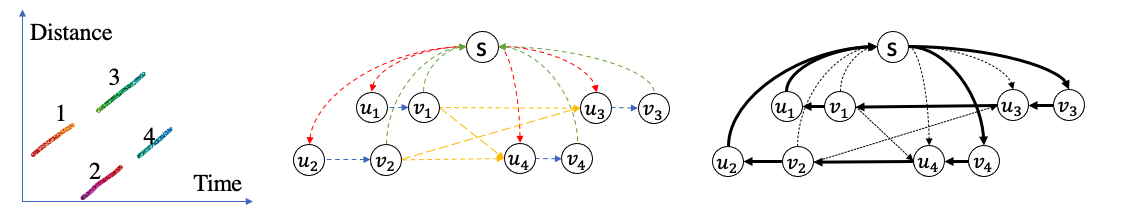}
    \caption{Left: fragments in time-space coordinates. In this example $\phi_1$ and $\phi_3$ should be associated, and $\phi_2$ and $\phi_4$ should be associated. The numbers indicate the order of last timestamp. Middle: fragments as a circulation graph. Red edges are the entering edges with cost $c_i^{en}$; blue edges are inclusion edges with cost $c_i$; green ones are exiting edges with cost $c_i^{ex}$ and yellow ones are transition edges with cost $c_{ij}$. Right: the residual graph after running the negative cycle canceling algorithm to obtain the min-cost circulation. The residual edges that carry the min-cost circulation are highlighted in bold. The fragment association assignment can be obtained by tracing along the bold edges.}
    \label{fig:mot_graph}
\end{figure*}

\subsection{Equivalent MCC formulation}
In seminal work~\cite{zhang2008global}, it is shown that~\eqref{eq:mcf} has a natural graph interpretation, and solving for~\eqref{eq:mcf} is equivalent to solving the min-cost-flow of a tracklet graph, which has a polynomial solution~\cite{ford1956maximal}. Later in the work of~\cite{wang2020efficient}, it is proven that the min-cost-flow problem for MOT is equivalent to a min-cost-circulation problem on a slightly modified graph. Many efficient algorithms are developed to solve this problem~\cite{ahuja1988network,Sokkalingam2000polynomial,Klein1966APM,goldberg1989finding}, and simplification are made to further improve the algorithmic efficiency in the MOT context~\cite{zhang2008global,lenz2015followme,wang2019mussp}. 
The graph is constructed such that each fragment $\phi_i$ is represented as two nodes $u_i$ and $v_i$, with a directed edge $(u_i \rightarrow v_i)$ and a cost $\$(u_i \rightarrow v_i)=c_i$ indicating \textit{inclusion} of $\phi_i$; edges between two fragments $\phi_i$ and $\phi_j$ are represented as $(v_i \rightarrow u_j)$, with the cost $\$(v_i \rightarrow u_j)=c_{ij}$ related to the likelihood of $\phi_j$ following $\phi_i$. The edge direction implies the sequential order between fragments. Furthermore, the graph has
a dummy node $s$ that has an incident edge to every $u$, and every $v$ directs back to $s$. The resulting graph is therefore a directed circulation graph, see Figure~\ref{fig:mot_graph}. We denote this circulation graph as $G(V,E)$, with node set $V$ and edge set $E$. Each edge $e:=(u,v) \in E$ has a unit capacity $r(e)=1$, a cost $\$(e)$ and a binary flow $f(e)\in \{0,1\}$. The data association problem can be formulated as finding a set of non-overlapping circulations $f$ on $G$ with the lowest total cost. The total cost of the circulations is $\sum_{e \in f} \$(e) f(e)$.



\subsection{Negative cycle canceling}
One efficient algorithm is the negative cycle canceling algorithm (NCC) proposed by Klein~\cite{Klein1966APM} and later on optimized by Goldberg et al.~\cite{goldberg1989finding,Sokkalingam2000polynomial}, based on the Ford-Fulkerson's method for incremental improvement. To understand the algorithm we first recall the definition of an important concept -- a residual graph $G_r$: 

\begin{definition}
\label{def: residual_graph}
The residual graph $G_r(V, E_r)$ for the original directed graph $G(V,E)$ with respect to a flow $f$ is generated by replacing each edge $e=(u \rightarrow v)\in E$ by two residual edges $e^{\prime}=(u\rightarrow v) \in E_r$ and $e_r =(v\rightarrow u) \in E_r$, with cost $\$(e^{\prime})=\$(e) $ and residual capacity $r(e^{\prime})=r(e)-f(e)$, while $\$(e_r)=-\$(e) $ and $r(e_r)=f(e) $.
\end{definition}
In the context of MOT graph as shown in Figure~\ref{fig:mot_graph}, the construction of residual graph can be simplified. The edges in the flow of the original graph simply needs to be reversed and costs on the edges negated, to form the corresponding residual graph.

The idea of NCC is to repeatedly find a cycle with negative cost in the residual graph $G_r$ and push flow through the cycles. The algorithm terminates when no more negative cycles can be found (optimality condition). We direct interested readers the above reference for the details and proof of correctness of this algorithm, and only provide an outline in Algorithm~\ref{alg:ncc}. 

First, a circulation graph $G(V,E)$ is constructed from the set of fragments $\Phi$ (ConstructTrackletGraph) and we iteratively look for a negative cycle in $G_r$ based on, for example, Bellman-Ford algorithm. If such cycle exists, then update the residual graph according to Definition~\ref{def: residual_graph} (PushFlow). When the iteration stops (no more negative cycle can be found), the assignment, or the trajectories, can be extracted by traversing along all the cycles through the residual edges in $G_r$ (FlowToTrajectories). 

\begin{figure*}
    \centering
    \includegraphics[width=\textwidth]{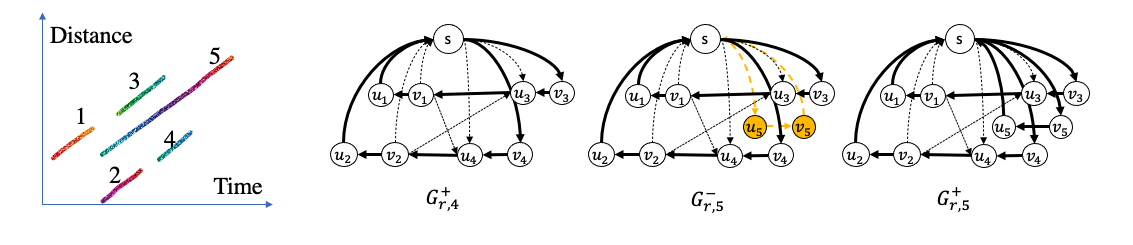}
    \caption{Scenario 1: the new fragment ($\phi_5$) starts a new trajectory. The residual graph from the previous iteration is $G_{r,4}^+$, with the min-cost circulation highlighed in bold. $G_{r,5}^-$ is obtained by AddNode($G_{r,4}^+$, $\phi_5$), with the added nodes highlighted in yellow. The min-cost cycle $\Gamma$ in $G_{r,5}^-$ is colored yellow. Finally, $G_{r,5}^+$ is obtained by PushFlow($G_{r,5}^-$, $\Gamma$).}
    \label{fig:ncc_s1}
    \centering
    \includegraphics[width=\textwidth]{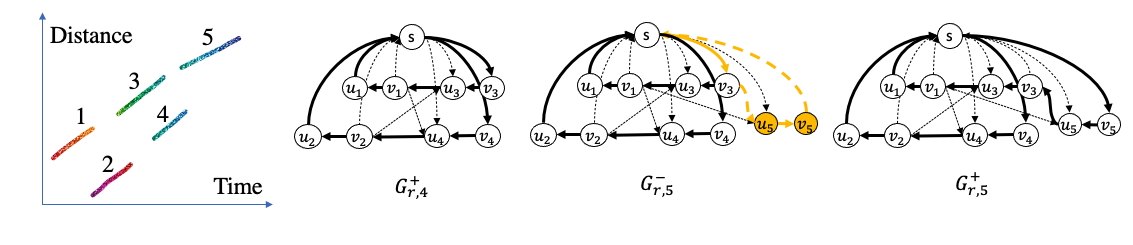}
    \caption{Scenario 2: the new fragment is connected to the tail of an existing trajectory. In this example the new node $u_5$ has candidate connections to $v_1$ and $v_3$ based on the motion model described in~\cite{wang2022automatic}. The min-cost cycle in this scenario also includes the post-node of $\phi_3$, $v_3$, which means that $\phi_5$ succeeds $\phi_3$ as the new tail of this trajectory.}
    \label{fig:ncc_s2}
    \centering
    \includegraphics[width=\textwidth]{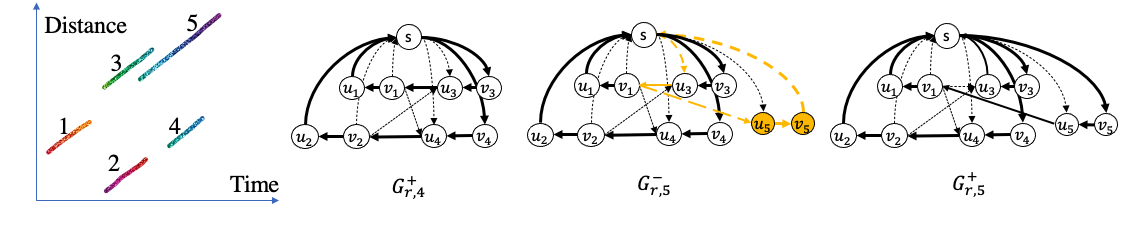}
    \caption{Scenario 3: the new fragment breaks an existing trajectory. In this case, the min-cost cycle contains the pre-node of $\phi_3$ and the post-node of $\phi_1$, meaning $\phi_5$ has a higher tendency to be a continuation of $\phi_1$ and $\phi_3$ is siloed.}
    \label{fig:ncc_s3}
\end{figure*}

  Note that the NCC algorithm guarantees feasibility at each iteration because every node is balanced (inflow equals outflow). The algorithm terminates immediately if no feasible flow can be found. The proof for correctness of the general NCC algorithm is detailed in~\cite{ahuja1988network}. The central idea is based on the theorem of negative cycle optimality conditions:
  \begin{theorem} (Negative Cycle Optimality Conditions)
  \label{thm:ncc_optimality}
     A feasible circulation $f$ in $G$ is optimal if and only if the residual graph $G_r$ contains no negative-cost cycles.
  \end{theorem}

Additionally we mention the following useful lemma that is specific to the MOT context, and was proved in~\cite{wang2020efficient}:
\begin{lemma}
\label{lemma:distinct_cycle}
    A circulation with total integer flow amount $K$ can only be sent through K distinct cycles.
\end{lemma}

  Next we show an online extension of the NCC algorithm and provide proof for correctness based on this important theorem.

\begin{algorithm}
\caption{Negative cycle cancellation for min-cost-flow on a tracklet graph}
\label{alg:ncc}
\SetKwInput{KwData}{Input}
\SetKwInput{KwResult}{Result}
\SetAlgoLined
    \KwData{Set of fragments $\Phi=\{\phi_i\}$}
    \KwResult{Set of trajectories $T=\{\tau_i\}$}

\quad $G(V,E,C)\leftarrow$ ConstructCirculationGraph($\Phi$)

\quad $f \leftarrow 0$

\quad $G_r \leftarrow G$


\While {a negative-cost cycle $\Gamma$ exists in $G_r$}{
    // Update residual graph\\
    $G_r \leftarrow$ PushFlow($G_r, \Gamma$)
 }       
 $T \leftarrow$ FlowToTrajectories($G_r$)
\end{algorithm}


\section{Online Negative Cycle Canceling}
\label{sec:online_ncc}
The streaming data coming from modern sensing technologies necessitates an online and memory-bounded version of Algorithm~\ref{alg:ncc}. In other words, the tracking graph $G$ is dynamic: new fragments are added in order of time and older fragments are removed from the graph. In this section we introduce an online version of the NCC algorithm which can be applied to a dynamic graph for online MOT. 

\subsection{Algorithm}
A naive online extension of Algorithm~\ref{alg:ncc} is to construct a circulation graph for each newly added fragment $\phi_k$ from scratch and rerun the NCC algorithm. However, it is inefficient because the majority of the graph remains the same and the majority of the computation on the min-cost cycle is wasted. This opens opportunities for an online extension of the algorithm to minimize repeated calculations.




    
   

The proposed online algorithm is based on the assumption that fragments are added to the graph in the order of last timestamp, which is a reasonable assumption in practice as this is the order that fragments are generated from object tracking. The online algorithm proceeds by adding each fragment $\phi_k$ to the residual graph from the previous iteration $G_{r,k-1}^+$ one at a time, to obtain a new graph $G_{r,k}^-$ (AddNode($G_{r,k-1}^-, \phi_k$)). This step adds two nodes $u_k$ and $v_k$ to the graph along with edges $(s\rightarrow u_k), (u_k\rightarrow v_k), (v_k \rightarrow s)$ and possibly additional transition edges incident to $u_k$. Then, we search for the least-cost negative cycle $\Gamma$ in $G_{r,k}^-$ (FindMinCycle($G_{r,k}^-$)) and push flow through the cycle to obtain the updated residual graph $G_{r,k}^+$. When all the fragments are processed, we output the trajectories $T$ by tracing all the cycles in the final residual graph. It can be proved that pushing flow through $\Gamma$, $G_{r,k}^+$ contains the min-cost circulation because the flow is feasible and no further negative cycles can be found in $G_r^+$. We denote the residual graph after adding $\phi_k$ at iteration $k$ to be $G_{r,k}^+$. The algorithm is shown in Algorithm~\ref{alg:ncc_online}.

A visual illustration of how the online NCC algorithm works is shown in Figure~\ref{fig:ncc_s1}-\ref{fig:ncc_s3}. We show three scenarios for adding a new fragment $\phi_5$ to the current MOT result with 4 fragments (The MOT result up to $k=4$ is maintained by $G_{r,4}^+$, where $\phi_1$ and $\phi_3$ are temporarily associated, and $\phi_2$ and $\phi_4$ are temporarily associated).
\begin{algorithm}
\caption{Online NCC for MCC on a tracklet graph}
\label{alg:ncc_online}
\SetKwInput{KwData}{Input}
\SetKwInput{KwResult}{Result}
\SetAlgoLined
    \KwData{Set of fragments $\Phi=\{\phi_i\}$}
    \KwResult{Set of trajectories $T=\{\tau_i\}$}

\quad $f \leftarrow 0$

\quad $G_{r,0}^{+} \leftarrow \{s\}$

\quad $k \leftarrow 1$

\For {each $\phi_k$ \textrm{(ordered by last timestamp)}} {
    $G_{r,k}^- \leftarrow$ AddNode($G_{r,k-1}^+$ , $\phi_k$) 
    
    $\Gamma \leftarrow $ FindMinCycle($G_{r,k}^-$)
    
    $G_{r,k}^+ \leftarrow $ PushFlow($G_{r,k}^-$, $\Gamma$)
   
   $k \leftarrow k+1$
}
 $T \leftarrow$ FlowToTrajectories($G_{r,k}^+$)
\end{algorithm}

\subsection{Proof for correctness}
Next we prove the correctness of Algorithm~\ref{alg:ncc_online}. Recall the negative cycle optimality condition in Theorem~\ref{thm:ncc_optimality}, we need to prove the following lemma:
\begin{lemma}
The circulation in $G_{r,k}^+$ is optimal, i.e., there is no more negative cycles in $G_{r,k}^+$ for every $k$.
\end{lemma}
\begin{proof}
We prove by induction. The base case is $G_{r,0}^+$, which contains a single node $s$ and therefore has no circulation nor negative cycle. During the first iteration, $G_{r,1}^-$ has one cycle: $s\rightarrow u_1\rightarrow v_1 \rightarrow s$. If the cost of this cycle is positive, then $G_{r,1}^+ = G_{r,1}^-$ and no more negative cycle remains in $G_{r,1}^+$. Otherwise if the cost for this cycle is negative, $G_{r,1}^+$ is $G_{r,1}^-$ with all edges reversed and costs negated, therefore the only cycle $G_{r,1}^+$ has a positive cost. 

For the induction, we want to prove that given $G_{r,k-1}^+$ which has no negative cycle, $G_{r,k}^+$ remains optimal (no negative cycles) after pushing flow through the min-cost cycle $\Gamma$ in $G_{r,k}^-$. Note that if $\Gamma$ does not exist on $G_{r,k}^-$, i.e., $G_{r,k}^+=G_{r,k}^-$, then $G_{r,k}^+$ remains optimal. If $\Gamma$ exists, it is obvious that $\Gamma$ must contain the subpath $u_k \rightarrow v_k \rightarrow s$ (one of the three scenarios illustrated in Figure~\ref{fig:ncc_s1}-\ref{fig:ncc_s3}). We proceed the proof by contradiction. 
    
Suppose there exists a negative-cost cycle $\Delta$ in $G_{r,k}^+$. Let $\bar{\Gamma}$ be the residual cycle in $G_{r,k}^+$ obtained by reversing and negating the cost of $\Gamma$.
If $\bar{\Gamma}$ and $\Delta$ share no common edges, then $\Delta$ must not contain the subpath $u_k \rightarrow v_k \rightarrow s$, nor any incident edges to $u_k$. This is because any flow that goes through an incident edge to $u_k$ must come out through the edge $u_k \rightarrow v_k$ per the flow conservation constraint. Therefore if $\Delta$ exists it must have already existed in $G_{r,k-1}^+$, which contradicts the precondition that $G_{r,k-1}^+$ is optimal (no negative cycles).

On the other hand if there exists a subpath $\pi(u,v) \in \Gamma$ and the residual path $\pi(v,u) \in \Delta$, i.e., $\bar{\Gamma}$ and $\Delta$ share a common subpath $\pi(v,u)$, we can prove that $\Gamma$ is not the min-cost cycle in $G_{r,k}^-$. Let the subpath $\pi(u,v)$ and $\Gamma^{'}$ form the cycle $\Gamma$, and the subpath $\pi(v,u)$ and $\Delta^{'}$ form the cycle $\Delta$ (see Figure~\ref{fig:ncc_proof}). We have $\$(\pi(u,v))=-\$(\pi(v,u))$ given the definition of a residual graph, along with the assumptions that $\Gamma$ and $\Delta$ are negative cost:
\[\$(\Gamma)=\$(\pi(u,v)) + \$(\Gamma^{'}) < 0, \textrm{and}\] 
\[\$(\Delta)=\$(\Delta^{'}) - \$(\pi(v,u)) < 0,\]
We can get
\[\$(\Gamma^{'}) +\$(\Delta^{'}) =\$(\Gamma) +\$(\Delta),\] meaning that the cycle formed by $\Gamma^{'}$ and $\Delta^{'}$ in $G_{r,k}^-$ has a lower cost than either $\Gamma$ or $\Delta$. It contradicts the fact that $\Gamma$ is the least-cost cycle on $G_{r,k}^{-}$. 

\begin{figure}
    \centering
    \includegraphics[width=0.8\linewidth]{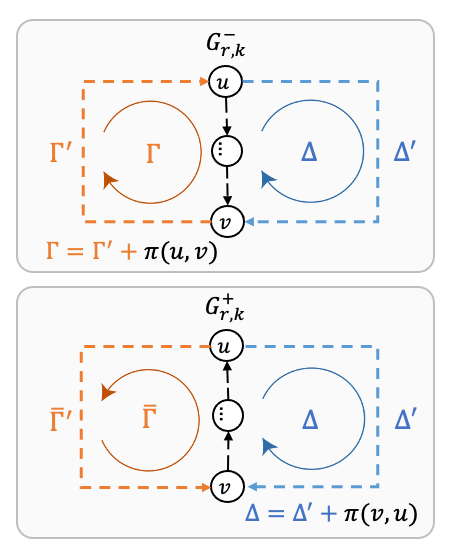}
    \caption{Proof that the larger cycle composed of $\Delta^{'}$ and $\Gamma^{'}$ is of lesser cost than $\Gamma$ in $G_{r,k}^-$.}
    \label{fig:ncc_proof}
\end{figure}
Therefore we proved that $G_{r,k}^{+}$ obtained from each iteration in Algorithm~\ref{alg:ncc_online} must be optimal.
\end{proof}

\subsection{Improvements}

We outline a few improvements on the runtime and memory of running the online NCC algorithm in practice.

\subsubsection{Runtime improvements}
The offline NCC algorithm has time complexity of $\boldsymbol{O}(|V||E|^2\log{|V|})$, given that the step of finding the minimum mean cycle (the cycle whose average cost per edge is smallest) takes $\boldsymbol{O}(|V||E|)$. The FindMinCycle step in the online NCC algorithm can be further improved based on the following result from the proof in the previous section:
\begin{corollary}
If a negative cycle exists on $G_{r,k}^-$ after adding the new fragment $\phi_k$, then it must contain a subpath $\pi(u_k,s) = (u_k \rightarrow v_k\rightarrow s)$.
\end{corollary}
This observation is helpful because we can limit our search for the min-cost cycle at each iteration to include this subpath. In order to find $\Gamma$, which has the cost of $\$(\Gamma) = \$(\pi(s,u_k))+\$(\pi(u_k,s))$, we simply need to find the shortest path from $s$ to $u_k$ in $G_{r,k}^-$ and check if $\$(\pi(s,u_k))+\$(\pi(u_k,s))<0$, as there is only one path for $\pi(u_k,s)$ and the cost of which is fixed. The step of FindMinCycle can be reduced to finding the single-source shortest path, which reduced the runtime to $\Theta(|E|+|V|\log|V|)$ at every iteration.



\subsubsection{Memory bound}
To limit the size of the graph at each iteration, we add a step CleanGraph($G_{r,k}^+, \tau$) to remove the trajectory (circulation) that is timed out at a customized time threshold $\tau$. Since all the fragments are added in order of time, we can simply check the tails of each trajectory, or the succeeding nodes to the dummy node $s$ at each residual graph $G_{r,k}^+$ for timeout. If timeout exceeds $\tau$, all the nodes along the entire circulation (except for $s$) can be safely removed from $G_{r,k}^+$. 

The removal of a circulation keeps the remaining flow in $G_{r,k}^+$ feasible because according to Lemma~\ref{lemma:distinct_cycle}, removing one cycle does not interfere with other cycles as no cycles have shared edges. The remaining of $G_{r,k}^+$ is still optimal because no negative cycle can be created in a subgraph of an optimal residual graph.

\begin{algorithm}
\caption{Memory-bounded online NCC}
\label{alg:ncc_online_mb}
\SetKwInput{KwData}{Input}
\SetKwInput{KwResult}{Result}
\SetAlgoLined
    \KwData{Set of fragments $\Phi=\{\phi_i\}$}
    \KwResult{Set of trajectories $T=\{\tau_i\}$}

\quad $f \leftarrow 0$

\quad $G_{r,0}^{+} \leftarrow \{s\}$

\quad $k \leftarrow 1$

\quad $\tau \leftarrow$ A time window 

\For {each $\phi_k$ \textrm{(ordered by last timestamp)}} {
    $G_{r,k}^- \leftarrow$ AddNode($G_{r,k-1}^+$ , $\phi_k$) 
    
    $\Gamma \leftarrow $ FindMinCycle($G_{r,k}^-$)
    
    $G_{r,k}^+ \leftarrow $ PushFlow($G_{r,k}^-$, $\Gamma$)

   $G_{r,k}^+ \leftarrow$  CleanGraph($G_{r,k}^+, \tau$)
   
   $k \leftarrow k+1$
}
 $T \leftarrow$ FlowToTrajectories($G_{r,k}^+$)
\end{algorithm}

\section{Experiments}
\label{sec:experiments}
In this section, we showcase the practical application of the online NCC algorithm through a numerical experiment focused on tracking vehicles on a highway. The experiment utilizes vehicle trajectory data, which consists of bounding boxes captured at a rate of 10Hz. The trajectory data is generated using TransModeler, a traffic microsimulation software, and then artificially segmented and degraded into fragments representing partial trajectories. The objective of this experiment is to demonstrate that by sequentially processing these fragments using the proposed online NCC algorithm, we can successfully recover the ground truth data association assignments. Additionally, we aim to illustrate that the algorithm effectively manages the graph size and computational time, thereby providing a bounded solution.

The tracking performance can be evaluated with standard MOT metrics specified in~\cite{bernardin2008evaluating,milan2016mot16,li2009learning,ristani2016performance}:
\begin{itemize}
\setlength{\itemsep}{0pt}
    \item \textit{Switches per GT}: total number of track switches per ground truth trajectory (target: 0).
    \item \textit{Fragments per GT}: total number of switches from tracked to not tracked per ground truth trajectory (target: 0).
\end{itemize}

\subsection{Experiment setup}

\begin{figure*}
    \centering
    \includegraphics[width=0.8\textwidth]{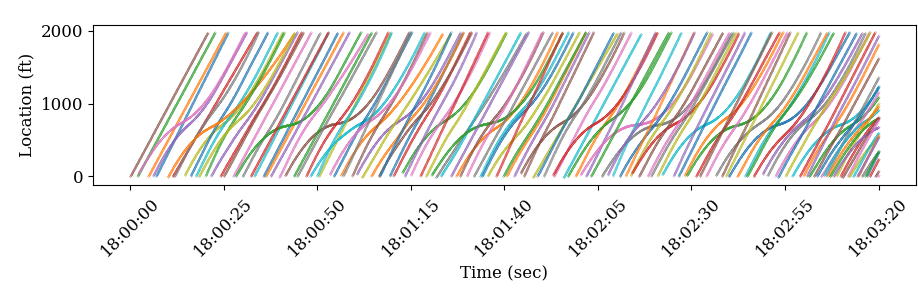}
    \caption{Time-space plot of the ground truth trajectories. Each object is represented as a distinct color.}
    \label{fig:SIM1}
    \centering
    \includegraphics[width=0.8\textwidth]{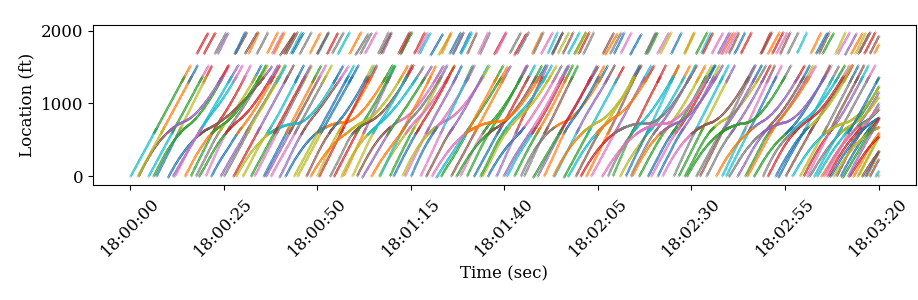}
    \caption{Fragments created from the simulated trajectory data. Each fragment is colored distinctly.}
    \label{fig:RAW1}
    \centering
    \includegraphics[width=0.8\textwidth]{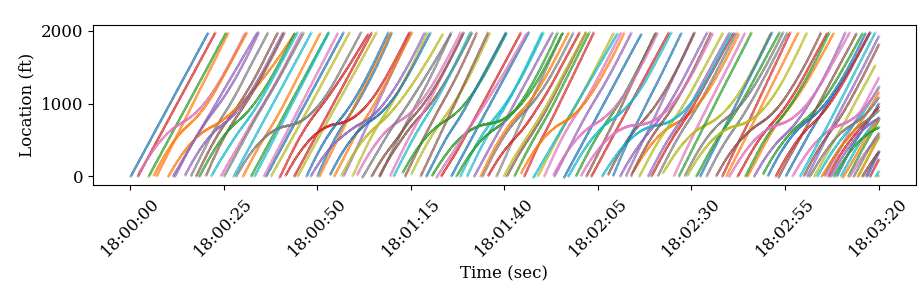}
    \caption{Data association result after online NCC, with additional data imputation and smoothing described in~\cite{wang2022automatic}.}
    \label{fig:REC1}
\end{figure*}

The simulation dataset we utilized consists of 137 vehicles moving along a 4-lane highway segment spanning a distance of 2000 ft. The simulation was conducted for a duration of 200 seconds, and lane-change occurs at random location and time throughout the simulation. A time-space diagram is shown in Figure~\ref{fig:SIM1}. Each trajectory is assigned a distinct color, allowing for easy differentiation.

To simulate the effects of conservative tracking commonly encountered in modern multi-camera systems, we introduced a mask that persists throughout the entire simulation timeframe. This mask covers the region from 1550 ft to 1700 ft and has a width of 150 ft. It mimics the discontinuity effect of a visual occlusion during the object tracking process, such as a highway overpass with vehicles passing underneath. Additionally, we incorporated two additional locations at 700 ft and 1400 ft where trajectories are fragmented. This behavior emulates the challenges faced when tracking across adjacent cameras with overlapping fields of view. As vehicles traverse between cameras, their IDs change, resulting in trajectory fragmentation. Furthermore, the fragments exhibit a slight overlap of approximately 100 ft, representing potential inaccuracies arising from camera misalignment or homography transformations. The time-space diagram of these fragments can be seen in Figure~\ref{fig:RAW1}.

The results for running the online NCC algorithm on the fragments are presented next.

\subsection{Results}
\begin{figure}[ht!]
    \centering
    \includegraphics[width=\linewidth]{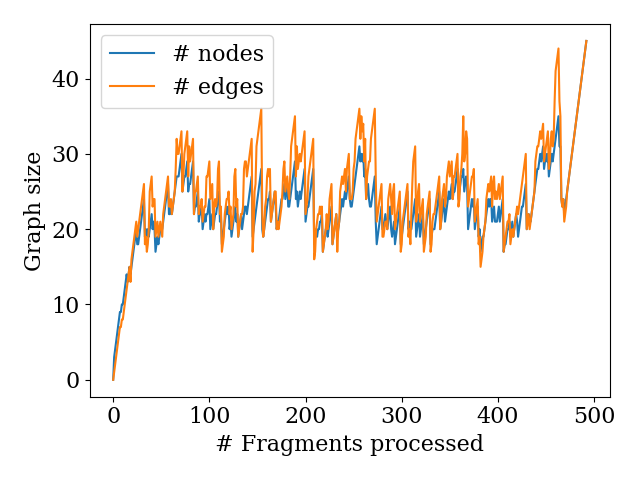}
    \caption{Graph size at each iteration of online NCC.}
    \label{fig:graph_size}
    \centering
    \includegraphics[width=\linewidth]{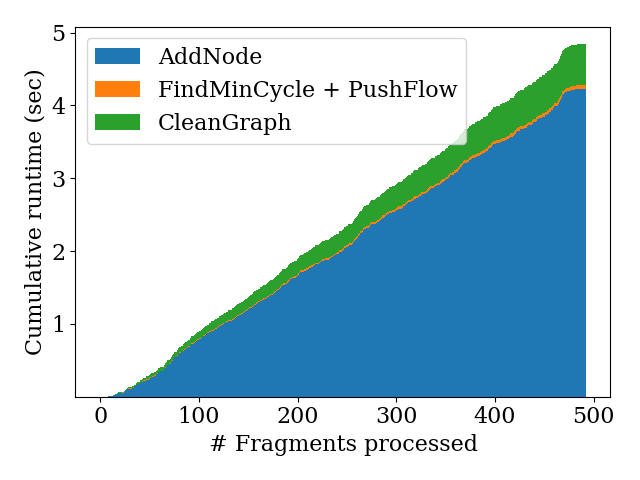}
    \caption{Cumulative runtime (sec) for each step during the online NCC.}
    \label{fig:runtime}
\end{figure}
First, we demonstrate that the online NCC algorithm applied to the fragments achieves 100\% accuracy. Table~\ref{tab:mota_1} illustrates that there are no mismatches between any fragments, as indicated by a count of 0 for both ``fragments per ground truth (GT)" and ``switches per GT". It verifies that all 137 ground truth objects are correctly recovered.

For this specific example, we assigned costs to each edge as follows: $c_i^{en}=c_i^{ex}=0$, $c_i=-10^{-6}$, and $c_{i,j}$ as a distance function that measures the motion similarity between two fragments, based on a probabilistic motion model (refer to equation (12) in \cite{wang2022automatic}). The rationale behind this choice is that there are no false positive fragments in this example; each fragment must be part of a real object's trajectory and thus no penalties on the entering and exiting edges. Additionally, a small reward is assigned for including each fragment through $c_i$. If for other settings false positive fragments are present, the corresponding costs on those edges need to be adjusted.

Next, we demonstrate that the graph size remains bounded during the online NCC procedure. We choose a window size of 5 seconds, meaning that when adding fragment $\phi_k$ with a last timestamp of $t_k$, all existing trajectories with a last timestamp older than $t_k-5$ are removed from the graph using the CleanGraph step. Figure~\ref{fig:graph_size} illustrates the relationship between the graph size and the number of fragments processed $k$. We observe that the number of nodes is maintained between 20-30, and the number of edges between 20-35, ensuring it remains ``memory-bounded". The graph size decreases each time when timed-out trajectories (nodes) are removed from the graph. Towards the end of the iteration, the size temporarily increases because fewer nodes are removed as the time cursor no longer advances forward.

The cumulative run-time for each process in the online NCC algorithm is shown in Figure~\ref{fig:runtime}. The total run-time of this example is approximately 5 seconds, averaging about 0.01 seconds per iteration. This time is well below the input rate and scales linearly with the number of fragments in the dataset. In this example, the majority of the run-time is consumed during the AddNode step, where the cost of every pair of fragments in the time window needs to be computed. Notably, the runtime of FindMinCycle and PushFlow combined constitutes only 1\% of the total computation time at each iteration ($10^{-4}$sec per iteration), which is the optimization target of the online NCC algorithm.

The numerical experiment setup for this study is deliberately simplistic, with the majority of vehicles traveling at free flow with a low lane-change rate, and the fragmentations occurring only at fixed locations. However, in real-world scenarios, fragmentations can occur randomly, and the lengths of tracks can vary significantly. Despite these complexities, the proposed algorithm remains versatile and does not impose any restrictions on the structure of the fragmentation or the location of missing tracks. For a more comprehensive evaluation of the proposed method, readers can refer to~\cite{wang2022automatic}, where a manually labeled dataset from the I-24 MOTION tracking system serves as the ground truth for benchmarking purposes. This extensive evaluation provides a more accurate assessment of the algorithm's performance under various real-world conditions and showcases its potential to handle more complex and diverse traffic scenarios.

\begin{table}
\centering
\begin{tabular}{@{}lccc@{}}
\toprule
Metrics / Statistics    & Ground truth & Fragments    &  After online NCC \\ \midrule

\# Distinct objects & 137 & 493 & 137\\
Fragments per GT $\downarrow$ & 0& 3.60 & 0 (-100\%)\\
Switches per GT $\downarrow$ &0 & 0 & 0 (-0\%)\\
\bottomrule
\end{tabular}
\caption{Evaluation results.}
\label{tab:mota_1}
\end{table}

\section{Conclusion}
\label{sec:conclusion}
We present an online extension of the well-known negative cycle canceling algorithm for solving the min-cost-circulation problem on a graph, which has the same global solution to the MOT problem. We provide proof for correctness of the algorithm, and demonstrate the application on a fragment association problem. The run-time analysis shows that the cumulative run-time is linear with respect to the input size, and is well below the real-time requirement. Additionally, the algorithm is shown to be memory-bounded, which can be suitable for settings with streaming input data.

\section*{Acknowledgment}
This study is based upon work supported by the National Science Foundation (NSF) under Grant No. 1837652, and the USDOT Dwight D. Eisenhower Fellowship program under Grant No. 693JJ322NF5201. 
\bibliographystyle{ieeetr}
\bibliography{refs}

\begin{thebibliography}{10}

\bibitem{chong2012graph}
C.-Y. Chong, ``Graph approaches for data association,'' in {\em 2012 15th
  international conference on information fusion}, pp.~1578--1585, IEEE, 2012.

\bibitem{castnnon2011multi}
G.~Castnn{\'o}n and L.~Finn, ``Multi-target tracklet stitching through network
  flows,'' in {\em 2011 Aerospace Conference}, pp.~1--7, IEEE, 2011.

\bibitem{zhang2008global}
L.~Zhang, Y.~Li, and R.~Nevatia, ``Global data association for multi-object
  tracking using network flows,'' in {\em 2008 IEEE Conference on Computer
  Vision and Pattern Recognition}, pp.~1--8, IEEE, 2008.

\bibitem{vyahhi2008tracking}
N.~Vyahhi, S.~Bakiras, P.~Kalnis, and G.~Ghinita, ``Tracking moving objects in
  anonymized trajectories,'' in {\em International Conference on Database and
  Expert Systems Applications}, pp.~158--171, Springer, 2008.

\bibitem{Rakai2022}
L.~Rakai, H.~Song, S.~Sun, W.~Zhang, and Y.~Yang, ``Data association in
  multiple object tracking: A survey of recent techniques,'' {\em Expert
  Systems with Applications}, vol.~192, p.~116300, 4 2022.

\bibitem{lenz2015followme}
P.~Lenz, A.~Geiger, and R.~Urtasun, ``Followme: Efficient online min-cost flow
  tracking with bounded memory and computation,'' in {\em Proceedings of the
  IEEE International Conference on Computer Vision}, pp.~4364--4372, 2015.

\bibitem{wang2022automatic}
Y.~Wang, D.~Gloudemans, Z.~N. Teoh, L.~Liu, G.~Zach{\'a}r, W.~Barbour, and
  D.~Work, ``Automatic vehicle trajectory data reconstruction at scale,'' {\em
  arXiv preprint arXiv:2212.07907}, 2022.

\bibitem{gloudemans202324}
D.~Gloudemans, Y.~Wang, J.~Ji, G.~Zachar, W.~Barbour, and D.~B. Work, ``I-24
  motion: An instrument for freeway traffic science,'' {\em arXiv preprint
  arXiv:2301.11198}, 2023.

\bibitem{wang2019mussp}
C.~Wang, Y.~Wang, Y.~Wang, C.-T. Wu, and G.~Yu, ``mussp: Efficient min-cost
  flow algorithm for multi-object tracking,'' {\em Advances in Neural
  Information Processing Systems}, vol.~32, 2019.

\bibitem{ford1956maximal}
L.~R. Ford and D.~R. Fulkerson, ``Maximal flow through a network,'' {\em
  Canadian journal of Mathematics}, vol.~8, pp.~399--404, 1956.

\bibitem{wang2020efficient}
C.~Wang, Y.~Wang, and G.~Yu, ``Efficient global multi-object tracking under
  minimum-cost circulation framework,'' {\em IEEE transactions on pattern
  analysis and machine intelligence}, 2020.

\bibitem{ahuja1988network}
R.~K. Ahuja, T.~L. Magnanti, and J.~B. Orlin, {\em Network flows}.
\newblock Cambridge, Mass.: Alfred P. Sloan School of Management,
  Massachusetts~…, 1988.

\bibitem{Sokkalingam2000polynomial}
P.~T. Sokkalingam, R.~K. Ahuja, and J.~B. Orlin, ``New polynomial-time
  cycle-canceling algorithms for minimum-cost flows,'' {\em Networks}, vol.~36,
  no.~1, pp.~53--63, 2000.

\bibitem{Klein1966APM}
M.~J. Klein, ``A primal method for minimal cost flows with applications to the
  assignment and transportation problems,'' {\em Management Science}, vol.~14,
  pp.~205--220, 1966.

\bibitem{goldberg1989finding}
A.~V. Goldberg and R.~E. Tarjan, ``Finding minimum-cost circulations by
  canceling negative cycles,'' {\em Journal of the ACM (JACM)}, vol.~36, no.~4,
  pp.~873--886, 1989.

\bibitem{bernardin2008evaluating}
K.~Bernardin and R.~Stiefelhagen, ``Evaluating multiple object tracking
  performance: the clear mot metrics,'' {\em EURASIP Journal on Image and Video
  Processing}, vol.~2008, pp.~1--10, 2008.

\bibitem{milan2016mot16}
A.~Milan, L.~Leal-Taix{\'e}, I.~Reid, S.~Roth, and K.~Schindler, ``Mot16: A
  benchmark for multi-object tracking,'' {\em arXiv preprint arXiv:1603.00831},
  2016.

\bibitem{li2009learning}
Y.~Li, C.~Huang, and R.~Nevatia, ``Learning to associate: Hybridboosted
  multi-target tracker for crowded scene,'' in {\em 2009 IEEE conference on
  computer vision and pattern recognition}, pp.~2953--2960, IEEE, 2009.

\bibitem{ristani2016performance}
E.~Ristani, F.~Solera, R.~Zou, R.~Cucchiara, and C.~Tomasi, ``Performance
  measures and a data set for multi-target, multi-camera tracking,'' in {\em
  European conference on computer vision}, pp.~17--35, Springer, 2016.

\end{thebibliography}
\end{document}